\documentclass[conference]{IEEEtran}
\usepackage{tikz}          
\usepackage{amsmath}
\usepackage[ruled]{algorithm2e}
\usepackage{graphicx}
\usepackage{booktabs}
\usepackage{latexsym}   
\usepackage{array}      
\usepackage{multirow}
\usepackage{subcaption}
\usepackage[noend]{algpseudocode}
\usepackage{threeparttable}
\usepackage{paralist}   
\usepackage{xspace}
\usepackage{color}
\usepackage{adjustbox}
\usepackage{balance}
\usepackage{wasysym}
\usepackage{rotating}   
\usepackage{listings}
\usepackage{tabu}
\usepackage{pifont}
\usepackage{framed}
\let\labelindent\relax
\usepackage[shortlabels]{enumitem}
\setenumerate[1]{itemsep=0pt,partopsep=0pt,parsep=\parskip,topsep=2pt}
\setitemize[1]{itemsep=0pt,partopsep=0pt,parsep=\parskip,topsep=2pt}
\setdescription{itemsep=0pt,partopsep=0pt,parsep=\parskip,topsep=2pt}
\usepackage{multirow}
\usepackage{url}            
\usepackage{hyperref}

\usepackage[normalem]{ulem}

\usepackage{cleveref}

\usepackage{xpatch}
\usepackage[frozencache,cachedir=.]{minted}

\usepackage{ragged2e}

\usepackage{amssymb}
\usepackage{cite}

\newcommand{\blue}[1]{\textcolor[rgb]{0.00,0.00,1.00}{#1}}

\definecolor{wheat1}{rgb}{1.000000,0.905882,0.729412}
\definecolor{LightGray}{rgb}{0.827451,0.827451,0.827451}

\newcolumntype{a}{>{\columncolor{wheat1}}l}

\definecolor{mygreen}{rgb}{0,0.6,0}
\definecolor{mygray}{rgb}{0.5,0.5,0.5}
\definecolor{mymauve}{rgb}{0.58,0,0.82}
\definecolor{darkblue}{rgb}{0.0,0.0,0.6}
\definecolor{maroon}{RGB}{102, 0, 0}
\definecolor{Maroon}{cmyk}{0,0.87,0.68,0.32}
\definecolor{darkred}{RGB}{139, 0, 0}
\definecolor{forestgreen}{RGB}{34, 139, 34}

\lstdefinelanguage{XML}
{
  basicstyle=\ttfamily\small,   
  morestring=[b]",
  moredelim=[s][\color{darkblue}]{<}{\ },
  moredelim=[s][\color{darkblue}]{</}{>},
  moredelim=[l][\color{darkblue}]{/>},
  moredelim=[l][\color{darkblue}]{>},
  morecomment=[s]{<?}{?>},
  morecomment=[s]{<!--}{-->},
  stringstyle=\color{darkred},
  identifierstyle=\color{mymauve}
}

\lstdefinestyle{customJava}{
  breaklines=true,
  keepspaces=true,
  frame=single,
  language=Java,
  showstringspaces=false,
  basicstyle=\footnotesize\ttfamily,
  keywordstyle=\color{blue},
  otherkeywords={+, getIntent},
  numbers=left,
  numbersep=5pt,
  numberstyle=\scriptsize\color{black},
  rulecolor=\color{black},
  stepnumber=1,
  tabsize=2,
  commentstyle=\itshape\color{green!40!black},
  stringstyle=\color{orange},
  emph=[1]  
  {
        do,
        try,
        new,
        catch,
        while,
        SecProvider,
        SecReceiver,
        SecService,
        SecActivity,
        SecSink,
  },
  emphstyle=[1]{\color{darkred}},
  emph=[2]  
  {
        @Override,
  },
  emphstyle=[2]{\color{purple!40!black}},
  belowskip=-1em, 
}

\newif\ifANNOYMIZE
\ANNOYMIZEtrue

\newif\ifACM
\ACMfalse 

\ifACM
\fi

\ifACM
\newcommand{\myfig}{Figure\xspace}
\else
\newcommand{\myfig}{Fig.\xspace}
\fi

\ifACM
\newcommand{\mysec}{\S}
\else
\newcommand{\mysec}{\S}
\fi

\newcommand{\code}[1]{{\fontfamily{cmtt}\fontseries{m}\fontshape{n}\selectfont\small{#1}}}

\definecolor{cadmiumgreen}{rgb}{0.0, 0.42, 0.24}

\newcommand\y{$\checkmark$\xspace}
\newcommand\x{\textcolor[rgb]{1.00,0.00,0.00}{$\times$}\xspace}

\newcommand{\name}{PropertyGPT\xspace} 
\newcommand{\tool}{PropertyGPT\xspace} 
\newsavebox{\bigimage} 

\newcommand{\CNumber}{23\xspace}

\usepackage{array}

\newcolumntype{L}[1]{>{\raggedright\let\newline\\\arraybackslash\hspace{0pt}}m{#1}}
\newcolumntype{C}[1]{>{\centering\let\newline\\\arraybackslash\hspace{0pt}}m{#1}}
\newcolumntype{R}[1]{>{\raggedleft\let\newline\\\arraybackslash\hspace{0pt}}m{#1}}
\newcommand{\website}{\url{https://github.com/Pr0pertyGPT/PropertyGPT}\xspace}

\newcommand{\etal}{\textit{et al.}\xspace}
\newcommand{\Certora}{Certora\xspace}
\newcommand{\prepost}{pre-/post-conditions\xspace}
\makeatletter
\chardef\TPT@@@asteriskcatcode=\catcode`*
\catcode`*=11
\xpatchcmd{\threeparttable}
  {\TPT@hookin{tabular}}
  {\TPT@hookin{tabular}\TPT@hookin{tabu}}
  {}{}
\catcode`*=\TPT@@@asteriskcatcode
\makeatother
\usepackage{tcolorbox}
\tcbuselibrary{breakable, skins}
\tcbset{%
  label begin/.style={label={#1}},
  label end/.style={after upper=\label{#1}}
}
\newtcolorbox[%
auto counter]{mybox}[2][]{%
  enhanced jigsaw,
  breakable,
  #1}

\begin{document}

\title{\name: LLM-driven Formal Verification of Smart Contracts through Retrieval-Augmented Property Generation}

\author{
\IEEEauthorblockN{Ye Liu$^1$\thanks{This work was done while Ye Liu was a student at Nanyang Technological University.}, Yue Xue$^2$\textsuperscript{\textdagger}\thanks{\textdagger Yue Xue and Ye Liu are the co-first authors.}, Daoyuan Wu$^3$$^*$\thanks{$^*$Daoyuan Wu is the corresponding author.}, Yuqiang Sun$^4$, Yi Li$^4$, Miaolei Shi$^2$, and Yang Liu$^{4,5}$}
\IEEEauthorblockA{$^1$Singapore Management University\\
$^2$MetaTrust Labs\\
$^3$The Hong Kong University of Science and Technology\\
$^4$Nanyang Technological University\\
$^5$China-Singapore International Joint Research Institute (CSIJRI)\\
}
}

\IEEEoverridecommandlockouts
\makeatletter\def\@IEEEpubidpullup{6.5\baselineskip}\makeatother
\IEEEpubid{\parbox{\columnwidth}{
		Network and Distributed System Security (NDSS) Symposium 2025\\
		24-28 February 2025, San Diego, CA, USA\\
		ISBN 979-8-9894372-8-3\\
		https://dx.doi.org/10.14722/ndss.2025.241357\\
		www.ndss-symposium.org
}
\hspace{\columnsep}\makebox[\columnwidth]{}}

\maketitle

\begin{abstract}
Formal verification is a technique that can prove the correctness of a system with respect to a certain specification or property.
It is especially valuable for security-sensitive smart contracts that manage billions in cryptocurrency assets.
    Although existing research has developed various static verification tools (or provers) for smart contracts, a key missing component is the \textit{automated} generation of \textit{comprehensive} properties, including invariants, pre-/post-conditions, and rules.
Hence, industry-leading players like Certora have to rely on their own or crowdsourced experts to manually write properties case by case.

With recent advances in large language models (LLMs), this paper explores the potential of leveraging state-of-the-art LLMs, such as GPT-4, to transfer existing human-written properties (e.g., those from Certora auditing reports) and automatically generate customized properties for unknown code.  
To this end, we embed existing properties into a vector database and retrieve a reference property for LLM-based in-context learning to generate a new property for a given code.
While this basic process is relatively straightforward, ensuring that the generated properties are (i) \textit{compilable}, (ii) \textit{appropriate}, and (iii) \textit{verifiable} presents challenges.
To address (i), we use the compilation and static analysis feedback as an external oracle to guide LLMs in iteratively revising the generated properties.
For (ii), we consider multiple dimensions of similarity to rank the properties and employ a weighted algorithm to identify the top-K properties as the final result. 
For (iii), we design a dedicated prover to formally verify the correctness of the generated properties.
We have implemented these strategies into a novel LLM-based property generation tool called \name. 
Our experiments show that \name can generate comprehensive and high-quality properties, achieving an 80\% recall compared to the ground truth.
It successfully detected 26 CVEs/attack incidents out of 37 tested and also uncovered 12 zero-day vulnerabilities, leading to \${8,256} in bug bounty rewards.


\end{abstract}

\section{Introduction}
\label{intro}

Smart contracts are transaction-driven programs deployed and executed on blockchain platforms, automating the execution of digital agreements among users.
Most smart contracts are written in Turing-complete programming languages, such as Solidity~\cite{solidity}, and have been widely adopted on popular blockchain platforms like Ethereum~\cite{Ethereum} and BSC~\cite{BSC}.
Smart contracts are extensively used in decentralized applications such as DeFi~\cite{kim2021survey} and NFTs~\cite{arora2022smart}.
However, they are susceptible to various types of attacks, including integer overflow~\cite{tan2022soltype}, re-entrancy~\cite{rodler2018sereum}, front-running~\cite{sun2024gptscan}, and access control vulnerabilities~\cite{SPCon22, SoMo23}.
These vulnerabilities primarily arise from loopholes in smart contracts due to programming errors, incorrect implementations, and logical bugs~\cite{zhang2023demystifying}.

Formal verification is one of the most advanced approaches to identify contract loopholes by performing a comprehensive examination with different kinds of specifications.
To perform formal verification, it is necessary to generate customized formal specifications for different smart contracts.
Formal specifications for smart contracts usually include temporal logic properties and Hoare logic properties, as surveyed in~\cite{tolmach2021survey}.
\textit{Invariants} are the most common contract specification, stating a property that holds for any contract execution, followed by \textit{function \prepost} for particular functional usage, as well as \textit{rules} that cover cross-function properties.
In most cases, temporal logic properties can be transformed into Hoare properties that could be instrumented into smart contract code~\cite{permenev2020verx}.
Hence, existing works typically use Hoare-style specifications for vulnerability detection~\cite{echidna, Wang2020OSD}, inconsistency detection~\cite{chen2019tokenscope}, and correctness validation~\cite{wang2018formal, permenev2020verx}.

Despite the promise of formal verification in enhancing the security and reliability of smart contracts, one notable challenge remains: the community still lacks the automated generation of comprehensive properties for effective formal verification of smart contracts.
While several works have attempted this, they have not yet achieved the ultimate goal of automatically generating necessary properties, including invariants, \prepost, and rules, for an unknown contract code.
For example, InvCon~\cite{Liu2022Invcon, liu2024automated} can dynamically infer \textit{likely} contract invariants and function \prepost, but it requires historical transaction information.
Likewise, Cider~\cite{liu2022learning} and SmartInv~\cite{wang2024smartinv} employed a machine-learning-based approach to generate specifications through the training-and-inference paradigm, but only for the \textit{invariant} properties.
As a result, industry-leading players like Certora~\cite{Certora} have to rely on their own or crowdsourced experts~\cite{CertoraContest} to manually write properties case by case, which hinders the effective formal verification of smart contracts on a large scale.

In this paper, we explore how recent advances in large language models (LLMs)
could enable automated generation of comprehensive smart contract properties.
Given LLMs' strong capability for in-context learning (see background in \mysec\ref{sec:LLMICL}), we try to achieve effective transfer learning from existing human-written properties to customized properties for unknown code.
More specifically, we embed existing properties into a vector database and retrieve a reference property for LLM-based in-context learning to generate a new property for a given code.
In this way, we can generate diverse types of properties as long as there are existing samples for each type in the collected vector database.
Moreover, compared to the training-and-inference paradigm mentioned above~\cite{liu2022learning, wang2024smartinv}, our approach does not require the error-prone labeling process (we can directly use existing raw property results, such as those from Certora auditing reports), nor re-training when there is updated data.

While the basic property generation process is relatively straightforward, it is challenging to ensure that the retrieval-augmented properties are (i) \textit{compilable}, (ii) \textit{appropriate}, and (iii) \textit{verifiable}.
To address these challenges, we employ three novel designs and implement them in an LLM-driven system called \name.
\textit{First}, we use compilation and static analysis feedback as an external oracle to guide LLMs in iteratively revising the generated properties.
\textit{Second}, we consider multiple dimensions of similarity to rank the properties and find a balanced metric for all these dimensions. The resulting weighted algorithm thus identifies the top-K properties as the final result.
\textit{Third}, we design a dedicated prover to formally verify the correctness of the generated top-K properties.

To evaluate \name, we collected 623 human-written properties from 23 Certora projects.
We first split 90 of them as a ground-truth testing set and used the rest as reference properties.
We found that \name can cover 80\% equivalent properties in the ground truth as judged by human experts, with a reasonable precision of 64\%.
Note that the additional properties (FPs) produced by \name generally also hold, complementing the human-written ones.
We further used all 623 properties as a knowledge base to supply \name for detecting real-world CVEs and past attack incidents.
Our results showed that \name successfully detected 9 out of 13 CVEs and 17 out of 24 attack incidents.
Moreover, during this process, \name demonstrated sufficient generalizability in analyzing an entirely different dataset.
Furthermore, we ran \name on four real-world bounty projects to demonstrate its ability to find zero-day bugs.
\name successfully generated 22 bug findings, out of which 12 have been both confirmed and fixed, earning us a total of \$8,256 in bounty rewards.

\noindent
\textbf{Contributions.}
We summarize the contributions as follows:
\begin{itemize}
\item We proposed a novel LLM-based property generation tool, \name, to drive comprehensive formal verification for smart contracts, with the major step of retrieval-augmented property generation described in \mysec\ref{sec:design}.

\item To facilitate \name, we also designed a property specification language (PSL) for smart contracts (\mysec\ref{sec:lang}) and a dedicated prover for property verification (\mysec\ref{sec:prover}).

\item We conducted extensive experiments and ablation studies to evaluate \name in various real-world settings; see \mysec\ref{sec:implement} and \mysec\ref{sec:evaluate}.
\end{itemize}

\noindent
\textbf{Availability.}
The property dataset and raw experimental data are available at \website, while the prototype is being commercialized by our industry partner, MetaTrust Labs. A partially open-source version will be updated on the above GitHub link. 

\section{Preliminary}
\label{sec:LLMICL}

\noindent
\textbf{Large language models (LLMs)}, such as GPT-3.5~\cite{ouyang_training_2022} and CodeLLama~\cite{CodeLLama}, have been widely used in many natural language processing tasks, such as text generation, translation, and summarization.
GPT series models are trained on a large corpus of text data and have the potential to generate human-like text, while CodeLLama is a fine-tuned version of LLama~2~\cite{LLama2} on open-source code.
The LLMs are pre-trained on a large corpus of text data and then fine-tuned on specific tasks and these datasets usually contain code from different programming languages.
Additionally, the pre-trained LLMs have exercised its potential to revolutionize the traditional software tasks, e.g., code generation~\cite{KLEE}, repairing~\cite{Xia_Zhang_2023,LLMZeroShotRepair}, vulnerability detection~\cite{sun2024gptscan}.

\noindent
\textbf{In-context learning (ICL).}
Based on the pre-trained knowledge, LLMs could leverage existing human-written properties written with various specification languages.
Yet, due to the limitations of the pre-training data and the efforts needed for training, LLMs may not be able to include the real-time information.
To address this problem, in-context learning (ICL) mechanism have been proposed by offering LLMs with the ability to learn from the latest conversation or task context~\cite{CanICL,OnTheEffectOfPretrainingCorporaOnInContextLearningByALargeScaleLanguageModel}.
In essence, in-context learning is a specialized kind of few-shot learning~\cite{LLMFewShot}, basing itself on a few examples or a small amount of data to learn a new task.


Instead of using fine-tuning~\cite{wang2024smartinv}, in this work, we employ the in-context learning ability from the state-of-the-art GPT-4 model~\cite{achiam2023gpt} for retrieval-augmented property generation.

\begin{figure*}[t]
	\centering
	\includegraphics[width=0.9\linewidth]{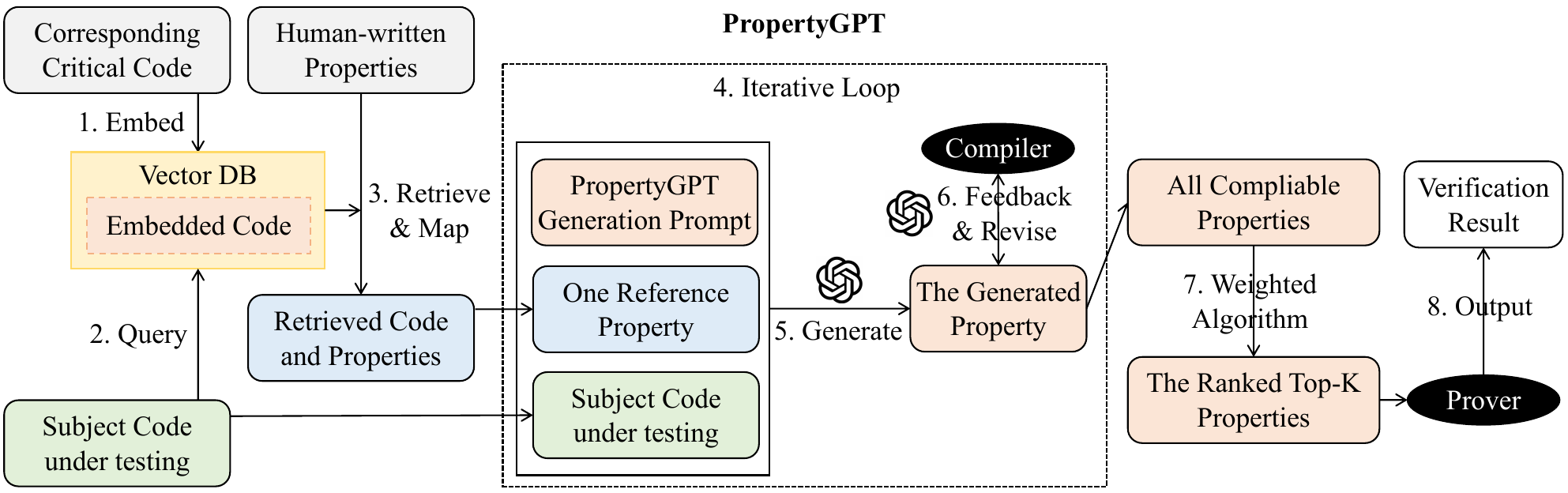}
    \caption{A high-level workflow of \name.}
	\label{fig:overview}
\end{figure*}

\section{\name Overview}
\label{sec:overview}

In this section, we present the overall design of \name, which leverages LLMs' ICL capability to transfer existing human-written properties and generate customized properties for formally verifying unknown code. 
At a high level, \name takes a piece of subject smart contract code as input and ultimately produces its corresponding properties along with the verification results.

As illustrated in \myfig~\ref{fig:overview}, \name consists of eight major steps:
\ding{172} \name first creates a vector database for reference properties by embedding their corresponding critical code. Note that the reference properties themselves will not be embedded because they are not the search key.
\ding{173} Given a piece of subject code under testing (typically one function), \name queries the vector database to \ding{174} retrieve all similar code within the threshold and map each code to their original reference properties.
\ding{175} All the reference properties are then tested with the subject code one by one in an iterative loop.
\ding{176} For each reference property, \name employs a generation prompt to generate a candidate property for the subject code.
\ding{177} This candidate property is then checked by the compiler for grammar, and if it is not grammatically correct, it will be further revised according to the compiler's feedback using a revising prompt.
\ding{178} Eventually, we obtain a list of compilable properties and rank them according to our weighted algorithm for the top-K appropriate properties.
\ding{179} These properties are finally formally verified by our prover, aiming to discover smart contract vulnerabilities.

To explore and understand the details of \name, we first introduce its property specification language in \mysec\ref{sec:lang}.
Following this, we describe the main process of LLM-based property generation and refinement in \mysec\ref{sec:design}.
Finally, we connect the generated properties with our dedicated formal prover in \mysec\ref{sec:prover} for property verification.

\section{Property Specification Language}
\label{sec:lang}

To bridge the gap between property generation in \mysec\ref{sec:design} and formal verification in \mysec\ref{sec:prover}, 
we propose an intermediate language in this section to specify the properties of smart contracts.

\myfig\ref{fig:psl} illustrates our property specification language (PSL), which extends the popular smart contract programming language Solidity. In Solidity, a smart contract (SC) consists of a group of state variables recording persistent program state and a list of public functions allowing user interactions. The symbol $\bowtie$ represents a set of arithmetic, comparison, or logical operators, namely $\{+,-,/, >, < ,== ,!=, >=, <=,\&, ||\}$. 
$bool\_expr\downharpoonright_{(v*, C*)}$ indicates boolean expressions involving state variables and constant values. 
PSL includes three kinds of properties with respect to different purposes. Invariants are properties that always hold true during contract execution and are defined over state variables; function \prepost are properties that can be expressed in Hoare triples $\{p*\}func\{q*\}$, checking whether parameters and the modification of state variables satisfy functionality requirements. Scenario-based properties, defined on restricted environments, can be implemented as different rules by enforcing varied assumptions and customized assertions.
It is worth noting that the current PSL prototype supports safety properties since they are more security-related compared with liveness properties.

\begin{figure}[t!]
	\small
	\begin{align*}
		\emph{v} \in \; StateVar& \quad \emph{tmp} \in \; TemporalVar \quad  \mathcal{C} \in\; Constant\\ \nonumber
		\emph{SC} &=  v*; func* \\ \nonumber 
		\emph{func}\in Function &= param*; stmt* \\ \nonumber
		\emph{expr} \in Expression &= tmp \;| v \;| \textbf{old}(v) | param\;| \mathcal{C} \;| expr \bowtie expr \\ \nonumber
		\emph{\textbf{Spec}(SC)} &=  inv*; \{p*\} func \{q*\}; rule*\nonumber \\
		\emph{inv} \in {Invariant} &= bool\_expr\downharpoonright_{(v*, C*)} \\ \nonumber
		\emph{p} \in {Precondition}  &= bool\_expr\downharpoonright_{(param, \textbf{old}(v), C)} \\ \nonumber
		\emph{q} \in {Postcondition}  &= bool\_expr\downharpoonright_{(param, v, C)} \\ \nonumber
		\emph{rule} \in {Rule} &=   \textbf{assume}(expr)* | func(expr*) | \textbf{assert}(expr)* \nonumber
	\end{align*}
    \caption{Property Specification Language (PSL).}
	\label{fig:psl}
\end{figure}


PSL brings several benefits to automate formal verification over existing works \cite{wang2018formal, Certora}.
VeriSol~\cite{wang2018formal} demands that assertion-based properties must be inserted into smart contract code. 
Certora's Verification Language (CVL)\footnote{https://docs.certora.com/en/latest/docs/cvl/index.html} is powered by a closed-source commercial verification tool, which restricts our ability to use CVL to build a self-contained pipeline for \name.
Additionally, the learning curve of CVL is quite steep since it requires not only knowledge about smart contracts but also several non-trivial techniques, such as using a \emph{hook} for data reading or writing, to handle the low-level execution model of smart contracts (an example of this limitation of CVL is provided in Appendix~\ref{sec:CVLexample}).
In contrast, writing or maintaining PSL specifications is easier because they share similar structures with the Solidity language~\cite{solidity}.
Note that the semantics of PSL will be illustrated in \mysec\ref{sec:verification}.

\section{Property Generation and Refinement}
\label{sec:design}

With the targeted PSL introduced in \mysec\ref{sec:lang}, our objective is to automatically generate properties written in PSL for the given code. The generated properties are the result of LLM-based transfer learning from existing human-written properties, which can be written in any specification language, not limited to PSL, such as CVL.

\name can achieve such powerful transfer learning fundamentally rooted in LLMs' capability for in-context learning (see \mysec\ref{sec:LLMICL}).
Nevertheless, we need to design a novel pipeline to facilitate this.
Our idea is to mimic the RAG (retrieval-augmented generation) process in the NLP~\cite{RAG} or code~\cite{LLM4Vuln24} domain, using the reference properties retrieved to augment the generation of new properties.
As previously illustrated in \myfig\ref{fig:overview}, we first detail how such retrieval-augmented property generation is conducted in \name in \mysec\ref{sec:rag}.
After that, \name iteratively revises the LLM-generated properties to fix their compilation errors in \mysec\ref{sec:revise}.
Furthermore, we design a weighted algorithm to help \name rank all compilable properties and obtain only the top-K appropriate properties for the prover's verification in \mysec\ref{sec:prover}.

\subsection{Retrieval-Augmented Property Generation}
\label{sec:rag}

Here we focus on the basic retrieval-augmented property generation that occurs from step \ding{173} to step \ding{176} in \myfig\ref{fig:overview}.
But the entire property generation process also includes property \ding{177} revising and \ding{178} ranking, which will be introduced in \mysec\ref{sec:revise} and \mysec\ref{sec:rank}, respectively.

\begin{figure}[t!]
\begin{tcolorbox}[title=Generation Prompt for Rule Properties]
Based on the rule code ([rule code]) and the code example ([code example]), generate corresponding rule code for [contract code to be tested].\\
1. Using the syntax style demonstrated in the provided code example, generate rule code. Focus on structural and syntactic aspects rather than replicating specific variable or function names from the example.\\
2. \$ is for a symbolic variable, such as \$varA for symbolic varA.\\
3. MUST NOT replicate specific variable or function names from the [code example].\\
4. MUST focus on the structural and syntactic aspects from the [code example].\\
5. When writing the rule code, closely follow the syntax and style from the provided example, focusing on its structural and syntactic essence rather than copying specific names.\\
6. The output MUST NOT contain any elements not predefined in the contract or function.

\tcbline
\text{[function code to be tested]}: \blue{\{func\_code\}}       \\
\text{[contract code to be tested]}: \blue{\{contract\_code\}}   \\
\text{[rule code]}: \blue{\{rule\_property\}}                    \\
\text{[code example]}: \blue{\{spec\_grammar\}}

\tcbline
The Output MUST be in the form of: \\
rule [name of rule]() \{\{logic of rule\}\}   \\
REMEMBER, ASSERT should not include an error message; just use the comparison operator directly.\\
REMEMBER, the rule must aim to test the function, not for another function.
\end{tcolorbox}

\caption{The prompt for generating rule properties.}
    \label{fig:ruleprompt}
\end{figure}

\begin{figure}[]
\begin{tcolorbox}[title=Generation Prompt for Function Invariants/Conditions]
Based on the following code ([condition code]), generate the corresponding precondition and postcondition code for [function code to be tested].

1. The basic syntax of preconditions and postconditions is in Solidity code format. \\
2. You can use the `\_\_old\_\_(xxx)` keyword if you need to reference the initial value of a variable. \\
3. You can directly use `xxxx==/!=/>/<` without `assert` or `require` to compare the value of the variable. \\
4. MUST NOT use `require` or `assert` for assertions; just use operator comparison directly. \\
5. MUST NOT use the ternary operator in the precondition and postcondition, but USE `if/else` expressions. \\
6. Exclude the event and implementation of the function itself, only output the precondition and postcondition of the function. \\
7. MUST NOT use any variables that I or the function have not defined, such as \_\_result\_\_, \_\_return\_\_, only follow the syntax I provide. \\
8. MUST NOT use `if/else` expressions in the precondition and postcondition, but USE the ternary operator. \\
9. MUST NOT INVOKE other functions or other undefined variables or non-state variables in the contract, only use the state variables in the {func\_name} itself. \\
10. Ignore and delete all conditions related to the return value.

\tcbline
\text{[function code to be tested]}: \blue{\{func\_code\}} \\
\text{[condition code]}: \blue{\{condition\_property\}}

\tcbline
The Output MUST be in the form of: \\
function \{func\_name\}\{\{
\begin{itemize}[label={},leftmargin=2em]
    \item precondition\{\{
    \begin{itemize}[label={},leftmargin=2em]
        \item Insert generated code here, ensuring it follows the syntax style of the example. 
    \end{itemize}
    \item \}\}
    \item postcondition\{\{
    \begin{itemize}[label={},leftmargin=2em]
        \item Insert generated code here, ensuring it follows the syntax style of the example. 
    \end{itemize}
    \item \}\}
\end{itemize}
\}\}
\end{tcolorbox}

\caption{The prompt for generating invariants/conditions.}
\label{fig:conditionprompt}
\end{figure}

\noindent
\textbf{\ding{172} Knowledge Preprocessing.}
One critical step in RAG-based systems is to first build a knowledge base, typically a vector database~\cite{LLM4Vuln24}.
In \name's scenario, we do not aim to extract ``knowledge'' from the existing human-written properties; instead, we use them as reference properties for LLMs' in-context learning.
Therefore, we directly use the raw information from human-written reference properties to construct our vector database.
As shown in \myfig\ref{fig:overview}, we embed the corresponding critical code of existing properties to build the search key used by RAG.
Note that the reference properties themselves will not be embedded because they cannot be queried against by the subject test code.

\noindent
\textbf{\ding{173}-\ding{174} Similar Example Retrieval.}
With the vector database, we can retrieve similar reference properties given the subject code to enable one-shot LLM learning in subsequent steps.
To do this, the subject code is also embedded in step \ding{173} and the dot product is calculated~\cite{FAISS} with all the vectors in the database.
The top similar code with the highest dot products are then retrieved, and their corresponding properties are returned as the result in step \ding{174}.
Here, we use a conservative code similarity threshold (e.g., 0.8) to limit the number of retrieved reference properties (typically 10 to 20 properties), which is acceptable because \name eventually uses a weighted algorithm in \mysec\ref{sec:rank} to rank only the top-K generated properties as the final result.

\noindent
\textbf{\ding{175}-\ding{176} In-context Learning.}
All the reference properties are then tested with the subject code one by one in an iterative loop.
For each reference property, \name employs a generation prompt to generate a candidate property for the subject code, using the reference property as a one-shot example.
Specifically, there are two types of generation prompts.
One is used to generate global, cross-function rule properties, as shown in \myfig\ref{fig:ruleprompt}, and the other is used to generate function-level \prepost, as illustrated in \myfig\ref{fig:conditionprompt}.
Note that here we omit the generation prompt template for contract-level invariants because they are usually in a simpler form and equivalent to the one-for-all \prepost for every public contract function.
Both prompt templates consist of three parts: the first part details the generation instructions, the second part lists the code and reference property, and the third part defines the output property format.
In particular, we supply a rule example in \myfig\ref{fig:ruleprompt} to help LLMs understand the grammar of our rule properties, while the grammar of invariants/conditions is directly specified using natural language instructions in \myfig\ref{fig:conditionprompt} as it is relatively simple.
Additionally, since rule properties are cross-function, we provide not only the function code but also the entire contract code in the prompt template shown in \myfig\ref{fig:ruleprompt}.

To determine which generation prompt should be used, \name leverages the type of the retrieved reference property.
If the reference property is a rule, \name uses the first type of prompt template to generate the property.
Otherwise, if the reference property is classified as {\prepost}, \name uses the second type of prompt template. 

\subsection{Revising Property to Fix Compilation Errors}
\label{sec:revise}

While the basic property generation process is relatively straightforward, one particular challenge is how to guarantee that the generated property is \textit{compilable}.
To address this challenge, we are inspired from~\cite{thakur2023autochip,kulsum2024case} and leverage the feedback from the compiler and static checking to iteratively revise the property until it is compilable or until it reaches the maximum number of attempts, as shown in step \ding{177}.

\begin{figure}[t!]
\begin{tcolorbox}[title=Common Prompt for Revising (Rule) Properties]
        Here is the rule I provided: \blue{\{spec\_res\}}. \\
        When this code is compiled with a solc-like program, an error occurs: \blue{\{error\_info\}}.

        \tcbline
Your task is to understand the rule I provided, fix the rule code, and correct the error within the rule. Refer to the contract code provided above. \\
Note, only modify the rule code; do not add other code. If the error is due to a non-existent variable, find feasible methods to reimplement it, or if it is not implementable, delete this line. \\
Here is the function code to be tested:
        \blue{\{func\_code\}} \\
Here is the contract code to be tested:
        \blue{\{contract\_code\}} \\
Provide me with the repaired rule code. The revised rule code must not be the same as the old rule code. \\
1. Using the syntax style demonstrated in the provided code example, generate a rule code. Focus on the structural and syntactic aspects rather than replicating specific variable or function names from the example. \\
2. \$ is for a symbolic variable, such as \$varA which symbolizes varA.

        \tcbline
        Rule Code Output MUST be in the form of:\\
        \texttt{rule [name of rule]()\{\{logic of rule\}\}}\\
        REMEMBER, ASSERT does not include an error message, just use the operator comparison directly. \\
        REMEMBER, the rule must aim to test the function [\blue{\{function\_name\}}], not for other functions. 
    \end{tcolorbox}

    \caption{The common prompt for revising (rule) properties.}
    \label{fig:commonreviseruleprompt}
\end{figure}

\begin{figure}[t!]
\begin{tcolorbox}[title=Special Prompt for Revising (Rule) Properties]
        Here is the knowledge rule you should learn from: \blue{\{knowledge\_rule\}}. \\
        Here is the rule I provided: \blue{\{spec\_res\}}. \\
        This rule lacks core function execution for: \blue{\{function\_name\}}. \\
        The contract code is: \blue{\{contract\_code\}}.

        \tcbline
        Your task is to understand the rule I provided, absorb the knowledge provided, and fix the rule code by adding the core function execution for: \blue{\{function\_name\}}. \\
        Here is the function code that needs to be tested: \blue{\{func\_code\}}. \\
Provide me with the revised rule code; the new rule code must not be the same as the old rule code. \\
1. Using the syntax style demonstrated in the provided code example, generate a rule code. Focus on the structural and syntactic aspects rather than replicating specific variable or function names from the example. \\
2. \$ is for a symbolic variable, such as \$varA for symbolic varA.

        \tcbline
        Rule Code Output MUST be in the form of:\\
        \texttt{rule [name of rule]()\{\{logic of rule\}\}}\\
        REMEMBER, ASSERT does not include an error message, just use the operator comparison directly. \\
        REMEMBER, the rule must aim to test the function [\blue{\{function\_name\}}], not for other functions.
    \end{tcolorbox}

    \caption{The special prompt for revising (rule) properties.}
    \label{fig:specialreviseruleprompt}
\end{figure}

\noindent
\textbf{Leveraging Compiler Feedback.}
Our PSL compiler (see its compilation details in Appendix~\ref{sec:compilationDetails}) provides compilation error information, including detailed error locations and reasons, if the property cannot be successfully compiled.
\name thus leverages this feedback to instruct LLMs to iteratively revise the property.
Specifically, we design and employ a common prompt template as shown in \myfig\ref{fig:commonreviseruleprompt} for revising rule properties; note that the prompt for revising {function \prepost} is similar and is therefore not shown here.
In this prompt, we ask LLMs to first understand the generated property code, identify and fix errors, maintain stylistic consistency throughout the process, and finally ensure that the revised rule code meets specific formatting requirements.
We set a threshold for the maximum number of attempts to avoid an endless loop.
In our experiment, as shown in \mysec\ref{sec:RQ3}, we found that 74\% of properties could be successfully compiled with no revisions (63\%) or with only one attempt, and {84\%} of all properties could be succesfully revised within five attempts.
This makes the iterative process manageable.

\noindent
\textbf{Employing Static Checking.}
However, we found that even when the compiler does not report any errors, it does not mean that the generated property is fully correct.
One notable issue is that LLM-generated properties could fail to include the target subject function, which renders the property meaningless.
To address this, we perform additional static checking for all compilable properties passed in the above step.
If \name identifies that the property is missing the target subject function, it employs a special prompt template as shown in \myfig\ref{fig:specialreviseruleprompt} (listing only the scenario for rule properties, similar to \myfig~\ref{fig:commonreviseruleprompt}).
It is generally similar to the common revising prompt but explicitly addresses the rule's lack of testing for the core function execution.
In this way, we not only guarantee that the generated property is grammatically correct but also functionally meaningful.

\subsection{Ranking the Top-K Appropriate Properties}
\label{sec:rank}

Another challenge is how to select the appropriate properties from all compilable properties as the final generation result.
To do so, we propose a weighted algorithm to rank all the resulting properties, as shown in step \ding{178}.
Specifically, we rank the properties based on the following four embedding-based metrics:
\begin{itemize}[leftmargin=2.5cm]
\item [$X_{raw}(f, g)$:] Similarity between contract code $f$ and $g$.
\item [$X_{summary}(f, g)$:] Similarity between high-level functionality summaries of code $f$ and $g$.
\item [$Y_{raw}(\phi_1, \phi_2)$:] Similarity between raw properties $\phi_1$ and $\phi_2$.
\item [$Y_{summary}(\phi_1, \phi_2)$:] Similarity between high-level property summaries for $\phi_1$ and $\phi_2$.
\end{itemize}
Note that we introduce $X_{summary}$ and $Y_{summary}$ to cope with that variety could exist for same-functionality code or same-semantic properties, where high-level natural language summaries are made by large language model for given code or properties.

Given an unknown code $f$, let $\phi_1$ be its property generated corresponding to reference code $g$ having property $\phi_2$,
we score  $\phi_1$ using a weighted algorithm as below.
 \begin{align}
 	Score(f, \phi_1) = \alpha * X_{raw}(f, g) + \beta * X_{summary}(f, g)  \nonumber \\  
 	+ \gamma * Y_{raw}(\phi_1, \phi_2) + \eta *  Y_{summary}(\phi_1, \phi_2) 
 \end{align}
 where $\alpha$, $\beta$, $\gamma$, $\eta$ are coefficients and $\alpha + \beta + \gamma + \eta = 1$.

To tune these coefficients, we train a linear regression model to approximate the actual property score $\hat{Score}(f, \phi_1)$, with details available in Appendix~\ref{sec:Coefficients}.
In this work, for simplicity, we consider $\hat{Score}(f, \phi_1) = Y_{summary}(\phi_1, \hat{\phi_1})$, where $\hat{\phi_1}$ is the corresponding ground truth property of $\phi_1$.
We have conducted a primitive experiment on 3,622 properties generated by \tool, and the results show that $\alpha$: 0.134, $\beta$: 0.556, $\gamma$: 0.141, and $\eta$: 0.168 are the optimal weights.

Consequently, properties with different scores are ranked in descending order, where we believe that properties with a higher rank are more likely to be important for the prover to verify.
 

\section{Property Verification}
\label{sec:prover}

\begin{figure}[t!]
	\includegraphics[width=\columnwidth]{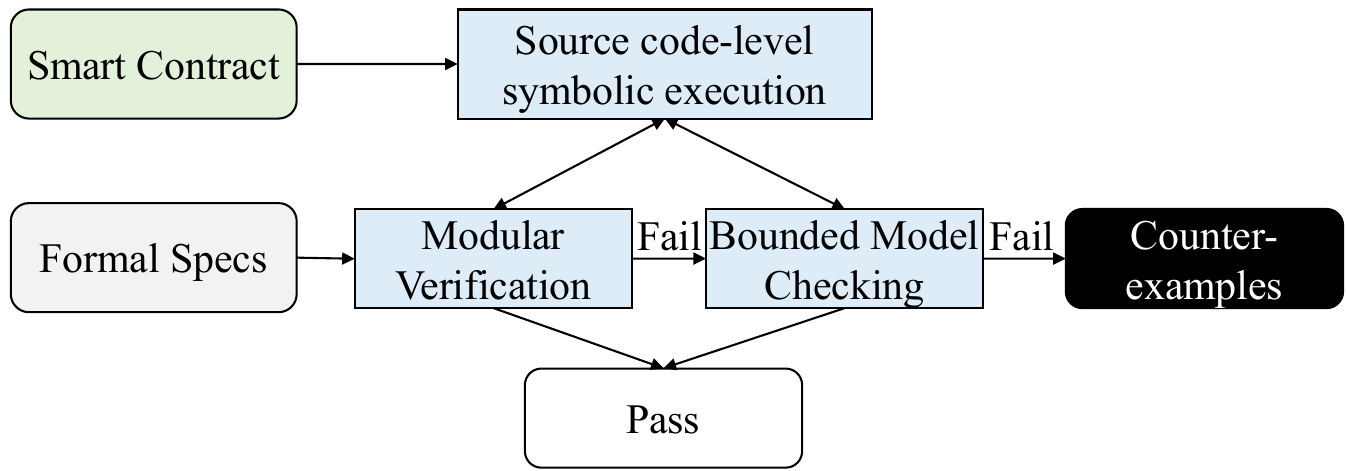}
	\caption{Workflow of Property Verification.}
	\label{fig:prover}
\end{figure}

The properties generated by \tool are not only \textit{compilable} and \textit{appropriate} but also \textit{verifiable}.
\myfig\ref{fig:prover} illustrates the workflow of our property verification process.
Our prover accepts smart contracts written in Solidity, along with their corresponding PSL specifications.
We employ forward symbolic execution to conduct a strongest postcondition analysis for each contract statement.
Subsequently, we perform modular verification to determine whether these formal specifications have been accurately implemented in the smart contract and can produce a proof if the properties hold.
In cases where the properties are violated, we use bounded model checking to verify whether the violated properties genuinely remain unfulfilled during contract execution.
Upon encountering counterexamples, we can confidently conclude that the properties indeed fail to hold, suggesting the presence of vulnerabilities in the smart contracts, which necessitate further manual verification.

\subsection{Modeling Smart Contract Execution}
\label{sec:strongest}

The runtime behaviors of smart contract execution rely on persistent contract states stored in the blockchain, transaction environment information, and specific contract statements to execute.
The persistent states are maintained in a group of contract state variables.
The transaction message includes the current block timestamp, the caller, the callee contract, and its called method.
Given a set of contract statements $\emph{S}=[s]$,
each statement execution can be modeled as a Hoare triple $\{\delta\} \;s\; \{\delta'\}$, where $\delta, \delta'$ indicate program states before and after executing $s$.
Unlike traditional programs, there is no crash in smart contract execution. 
Any unexpected behavior will cause a reversion of the smart contract transaction, leaving the contract state unchanged.
Such reversion behaviors may affect availability, e.g., denial of service, but generally do not pose a threat to smart contract safety and therefore we exclude them from our analysis.

\subsection{Verification Technique}
\label{sec:verification}
We employ the small-step operational semantics of smart contracts to formally verify invariants, function pre/post-conditions, and rules.
Readers can refer to KSolidity~\cite{jiao2020semantic} for detailed operational semantics.

\noindent
\textbf{Function Pre/post-condition Verification.}
Given a function $f$, let $p$ and $q$ be its precondition and postcondition to verify, respectively. Specification $\{p\}\; f\; \{q\}$ is provable if and only if the below predicate holds
\begin{align}
\forall \delta \in \Delta, sp(f, p \cap \delta) \implies q 
\end{align}
where $\Delta$ encompass all feasible contract states.

We elide verification details for contract invariants because invariants can be deemed as a variant of function-level specifications that hold for every contract function, with the same precondition and postcondition standing there.

\noindent
\textbf{Rule Verification.}
Given a rule-based property \emph{rule}, with user-given assumptions $\delta$ declared in \code{assume} statements and assertions $q$ declared in \code{assert} statements, \emph{rule} holds if and only if the below predicate holds
\begin{align}
		sp(rule, \delta) \implies q 
\end{align}
Note that $\delta$ may not always be feasible contract state.

We utilize source code-level symbolic execution to conduct strongest postcondition analysis for Solidity smart contracts.
Distinguishing ourselves from existing research~\cite{lin2022solsee}, our novel symbolic execution approach implements comprehensive small-step semantics, enabling automated analysis of real-world, complex smart contracts.
Although SolSee~\cite{lin2022solsee} has made strides in symbolic execution for Solidity smart contracts, it lacks support for certain critical features commonly used in smart contracts, such as the aggregated effect of intricate expressions and polymorphic handling during complex inheritance relationships.
We address these limitations by meticulously adhering to the practices of the Solidity compiler, ensuring precise semantics of complex expressions.
For instance, expressions are evaluated from left to right as specified by the compiler.
Furthermore, our approach accurately resolves polymorphism during both the compilation and execution stages of smart contracts.
To mimic actual contract execution, our symbolic execution approach maintains a comprehensive list of function signatures and revisits contract inheritance chains to determine the exact function implementation for ambiguous calls, such as \texttt{super().call()}, where \texttt{super} refers to an unknown parent contract.

To deal with the complexity of smart contracts, we implemented several over-approximation techniques to handle unknown or non-linear operation semantics.
First, the behaviors of function calls to on-chain smart contracts remain unknown at the verification stage, so we assume all on-chain calls succeed but make their return data symbolic to accommodate any possible outcome.
This approach is necessary because on-chain contracts may not be open source, and their inter-contract interactions can be overly complex, falling beyond the scope of our current research.
Second, non-linear native functions, such as \code{sha3}, which computes the hash value of a string, are challenging to model precisely.
To address this, we utilize uninterpreted functions~\cite{gange2016abstract} to capture their primary features, such as treating \code{sha3} as an injective function.

We employed modular verification and bounded model checking.
During modular verification, we lift all state constraints by making all state variables symbolic.
Correctness can be safely ensured when the specification properties hold accordingly.
Otherwise, we perform bounded model checking to systematically explore all feasible states to find counterexamples that violate the property being verified.
Any violated property and its counterexamples will be manually investigated to confirm the existence of vulnerabilities, similar to the approach used in SmartInv~\cite{wang2024smartinv}.
The depth of bounded model checking is capped at three by default, and we allow up to five loop iterations in the case of non-terminated execution.

\section{Implementation and Setup}
\label{sec:implement}

We implemented~\tool in around 3K lines of Python code for LLM-based property generation, and around 38K lines of C++ code for grammar support and verification of PSL property specifications. 
Additionally, for applying symbolic execution to smart contracts written in various versions of Solidity, we developed a converter to map smart contracts written in Solidity versions 0.6.x and 0.7.x into abstract syntax trees compatible with the latest Solidity version 0.8.x, where we have systematically investigated their syntax and semantic differences.
We use the Z3 solver, version 4.11.2, to discharge symbolic constraints for path feasibility checking and property satisfiability checking.

\subsection{Property Knowledge Collection}
To obtain high-quality human-written properties as the knowledge base for in-context learning, we systematically analyzed 61 audit reports from the \Certora platform, for which experts have written property specifications to facilitate formal analysis of smart contracts.
These audit reports were published from 2019 to 2023.
Through further investigation, we removed 38 projects whose contract code and raw properties were not available, and eventually, we collected 23 \Certora projects, including 623 human-written properties, which will be detailed in Table~\ref{tab:appendix} in Appendix~\ref{sec:appendixA}.
It is worth noting that for the selected projects, all human-written properties were collected, whether they represent violated or non-violated ones in their particular audited projects. 

\begin{figure}
	\includegraphics[width=\columnwidth]{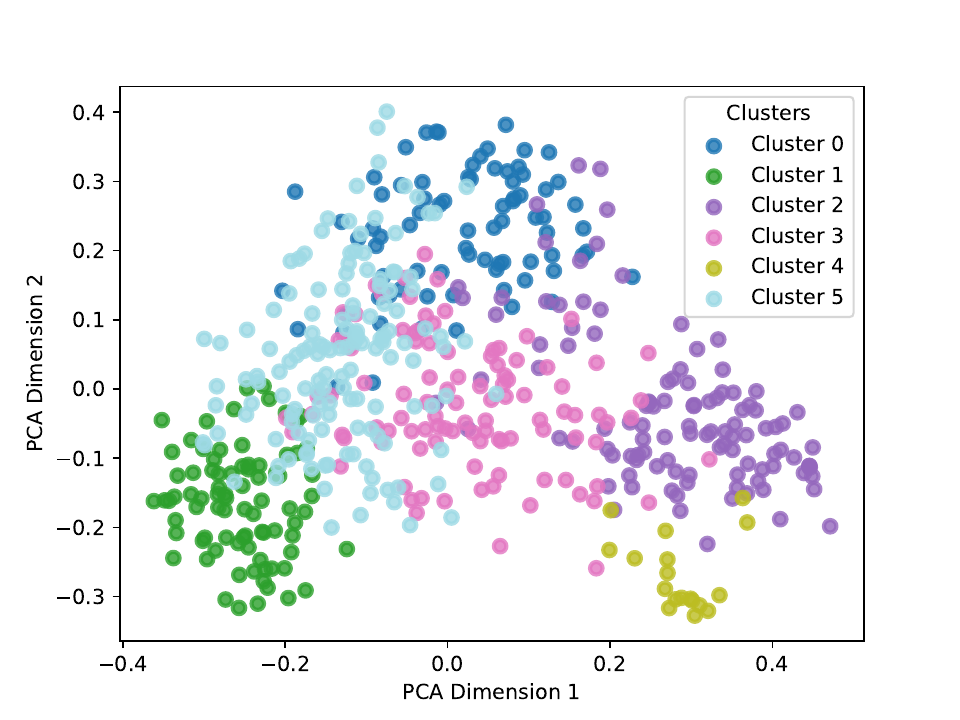}
    \caption{The property cluster distribution after two-dimension {PCA}~\cite{hasan2021review} reduction.}
	\label{fig:cluster}
\end{figure}

To study the characteristics of these properties, we employed the affinity propagation clustering algorithm~\cite{frey2007clustering} from the \code{sklearn}~\footnote{\url{https://scikit-learn.org/}} library to discern property categories, based on pairwise embedding similarity across the properties.
Specifically, we performed some preliminary experiments and found that this setting \textit{AffinityPropagation(damping=0.5, preference=-75, random\_state=5)} could establish a good result for property clustering.
\myfig\ref{fig:cluster} illustrates the distribution where six clusters are labeled with different colors.
However, it is clear that overlapping exists among clusters, especially for clusters \#3 and \#5.

Furthermore, we investigated all the clusters and have the following classifications of human-written properties as shown in Table~\ref{tab:property}.
There are six property categories as follows:
\begin{itemize}[leftmargin=*]
    \item \textit{DeFi}, involving the management of its essential protocol components including reserves, collateral, and liquidity pools;
    \item \textit{Token}, which is the cornerstone of entire DeFi ecosystems, specifying standard behaviors such as token balance and critical operations like transfer and minting;
    \item \textit{Arithmetic}, focusing on the correctness of numerical conversions and the consistency of asset splitting;
    \item \textit{Usability}, examining the validity of operations containing timestamp-based constraints (Temporal Use) and operations with contract state-based constraints (State-dependent Use);
    \item \textit{Governance}, which plays an important role in the management of decentralized applications, usually through a voting mechanism, concerning issues such as the transfer of voting power to a delegator and the double-voting problem;
    \item \textit{Security}, checking for the presence of common vulnerability types including front running and overflow.
\end{itemize}

\begin{table}
	\centering
	\caption{Characteristics of the collected \Certora properties.}
	\small
\resizebox{\columnwidth}{!}{
\begin{tabular}{llc}\toprule
	Category &Classification &Property Examples \\\midrule
	\multirow{3}{*}{DeFi } &Reserve & setReserveFactorIntegrity\\
	&Collateral & integrityOfisUsingAsCollateralAny\\
	&Liquidity & checkBurnAllLiquidity\\
	\midrule
	\multirow{4}{*}{Token } &Balance & total\_supply\_is\_sum\_of\_balances\\
	&Transfer/TransferFrom & transferBalanceIncreaseEffect\\
	&Mint/Burn & integrityMint, additiveBurn\\
	&Approve & approvedTokensAreTemporary\\
	\midrule
	\multirow{2}{*}{Arithmetic } &Numerical Conversion & toElasticAndToBaseAreInverse1down\\
	&Asset Splitting & moneyNotLostOrCreatedDuringSplit \\
	\midrule
	\multirow{2}{*}{Usability} &Temporal Use & timestamp\_constrains\_fromBlock\\
	&State-dependent Use & unsetPendingTransitionMethods\\
	\midrule
	\multirow{2}{*}{Governance} &Delegation & integrityDelegationWithSig\\
	&Voting & totalNonVotingGEAccountNonVoting\\
	\midrule
	\multirow{2}{*}{Security} &Front run & cannotFrontRunSplitTwoSameUsers\\
	&Overflow & integrityOfMulDivNoOverflow \\
	\bottomrule
\end{tabular}
}
	\label{tab:property}
\end{table}

\subsection{Experimental Setup}
We use the large language model GPT-4-turbo provided by OpenAI through its API \code{gpt-4-0125-preview}.
Regarding the model configuration, we adhere to the default settings where the temperature is 0.8, top-p is 1, frequency penalty and presence penalty are both 0, and the maximum response length is 2000.
Moreover, we calculate all embedding similarities using the pre-trained model \code{text-embedding-ada-002} from OpenAI.
For property generation, we cap revising attempts at nine to limit LLM usage for better economics.
All experiments were conducted on a Docker with Ubuntu 20.04 OS, an Intel Core Xeon 2.2 GHz processor, and 2GB RAM.


\section{Evaluation}
\label{sec:evaluate}

In this work, we aim to answer the following research questions (RQs):
\begin{itemize}[leftmargin=*]
\item RQ1: (\textbf{Property Generation}) How accurately does \tool generate properties for smart contracts?
\item RQ2: (\textbf{Vulnerability Detection}) How effectively does \tool discover smart contract vulnerabilities? Can \tool achieve state-of-the-art results?
\item RQ3: (\textbf{Generalizability}) Does \tool have sufficient generalizability to enable powerful transfer learning?
\item RQ4: (\textbf{Influencing Factors}) What factors influence the performance of \tool?
\item RQ5: (\textbf{Impact}) How well does \tool find zero-day vulnerabilities in real-world smart contract projects?
\end{itemize}

\noindent
\textbf{Methodology.}
To answer RQ1, we divide \Certora properties into a testing dataset and a ``training'' dataset as the knowledge base.
We instruct \name to generate properties for smart contracts in the testing dataset using smart contracts and their properties from the knowledge base.
We compare the resulting properties by \tool with ground-truth human-written ones to investigate its effectiveness.
Specifically, we randomly selected nine (40\%) \Certora projects as our testing dataset and then picked 10 properties for each project.
Consequently, our testing dataset includes 90 ground truth properties from nine projects.
During this experiment, \tool first extracts the subject function code where the ground truth properties are specified, and then queries the knowledge base to enable ICL to automate property generation.

To answer RQ2, we compare \tool with SmartInv~\cite{wang2024smartinv}, a concurrent work with ours, published recently in May 2024.
We contacted the authors to obtain a copy of their source code and benchmark\footnote{\url{https://github.com/sallywang147/attackDB}}, which includes 60 attack incident projects that have suffered significant losses.
Upon reviewing their benchmark, we identified several issues.
Among the listed cases, 2 are repeated, 9 lack public exploit transactions (e.g., sherlockYields), 2 are not open-sourced, and 2 have incomplete code.
Of the remaining cases, 11 are reentrancy attacks that could be easily remedied by adding the widely-used \code{nonReentrancy} modifier.
Furthermore, 8 cases involve price manipulation attacks, which may be impractical to identify using the simple invariant properties that SmartInv generated.
For example, Tolmach\etal\cite{tolmach2021formal} proposes a semi-automated formal composite analysis for DeFi protocols that detects such problems with fairness properties, while others use either runtime monitoring\cite{wu2021defiranger} to identify attack behavior or static analysis~\cite{kong2023defitainter} to flag vulnerable code with predefined patterns.
Through this deep analysis of their benchmark, we curated 24 attack incidents from the SmartInv benchmark for our evaluation.
Additionally, we compare \tool with state-of-the-art tools~\cite{sun2024gptscan, feist2019slither, mossberg2019manticore, Mythril} on well-known smart contract CVEs.
As of April, 2024, there are 577 smart contract CVEs, predominantly 477 integer overflows.
To avoid bias, we randomly selected 13 CVEs of different types: three integer overflow cases, three access control vulnerabilities, four other logic bugs, etc., details of which are provided in \Cref{tab:cve}.


\noindent
\textbf{Benchmarks.}
As shown in \Cref{tab:benchmark}, we evaluate the property generation process using \Certora audited projects and test the applicability of \tool in vulnerability detection using well-known CVEs and attack incident projects studied by SmartInv.
Additionally, RQ3 and RQ4 also use the corresponding benchmarks, the details of which will be elaborated later.

\begin{table}[t!]
	\caption{The evaluation benchmarks.}
	\small
	\begin{tabular}{ll}
		\toprule
		Benchmark & RQs \\
		\midrule
		\CNumber \Certora projects (623 properties; 90 for testing)  & RQ1, RQ4 \\ 
		13 CVEs + 24 projects from the SmartInv benchmark  &  RQ2, RQ3 \\ 
		\bottomrule 
	\end{tabular}
	\label{tab:benchmark}
\end{table}

\begin{table*}[t!]
	\centering
	\small
    \caption{The property generation results for 90 ground-truth properties from nine \Certora projects.}
		\begin{tabular}{lr|rrrrr|rrr}\toprule
			Project &\#Property (\Certora) &\#Property (ours) &TP &\#Hit &FN &FP &Recall &Precision &F1-score \\\midrule
			aave\_proof\_of\_reserve &3* &38 &25 &3 &0 &13 &\textbf{1.00} &0.66 &0.79 \\
			aave\_v3 &17 &61 &32 &15 &2 &29 &0.88 &0.52 &0.66 \\
			celo\_governance &10 &39 &29 &10 &0 &10 &\textbf{1.00} &0.74 &\textbf{0.85} \\
			furucombo &10 &23 &11 &7 &3 &12 &0.70 &0.48 &0.57 \\
			openzepplin &10 &2 &2 &1 &9 &0 &\textbf{0.10} &\textbf{1.00} &\textbf{0.18} \\
			opyn\_gamma\_protocol &10 &30 &14 &8 &2 &16 &0.80 &\textbf{0.47} &0.59 \\
			ousd &10 &100 &67 &10 &0 &33 &\textbf{1.00} &0.67 &0.80 \\
			radicle\_drips &10 &17 &9 &7 &3 &8 &0.70 &0.53 &0.60 \\
			sushi\_benttobox &10 &70 &49 &10 &0 &21 &\textbf{1.00} &0.70 &0.82 \\
			\midrule
			Average &10 &42 &26 &8 &2 &16 &0.80 &0.64 &0.71 \\
			\bottomrule
		\end{tabular}
		\begin{flushleft}
			* This project contains only three human-written properties, so we picked seven more from \code{aave\_v3}. Both are from the same institution.\\
		\end{flushleft}
	\label{tab:applicability}
\end{table*}

\subsection{RQ1: Property Generation}
\label{sec:RQ1}

We evaluated \tool on 90 ground-truth properties from nine \Certora projects to investigate the effectiveness of property generation.
\Cref{tab:applicability} shows \name's property generation results using the rest of \Certora properties as the knowledge base.
Note that this RQ only measures whether the generated properties match the ground-truth properties, where an FP indicates a property unmatched with the ground truth.

The first two columns show the project name and the number of properties written by \Certora experts, i.e., \#Property (\Certora).
The middle five columns list the number of properties generated by \tool, i.e., \#Property (ours), true positives that are equivalent to the ground-truth properties (TP), the number of ground-truth properties hit by the properties generated (\#Hit), the number of missed ground-truth properties (FN), as well as false positives (FP).
The last three columns are recall, precision, and F1-score metrics where $\text{recall} = \frac{\#Hit}{\#Hit+FN}$, $\text{precision} = \frac{TP}{TP+FP}$, and $\text{F1-score}=\frac{2 \times \text{recall} \times \text{precision}}{\text{recall}+\text{precision}}$.
Because \Certora properties are written in the proprietary \Certora Verification Language (CVL) that supports formal verification of smart contracts at the EVM bytecode level, while \tool uses PSL to facilitate property formulation and verification at the source code level,
there is currently no automated analysis tool available for equivalence checking between properties from these two distinct specification systems.
Therefore, two authors with 5 years of research and auditing experience independently examined the equivalence between the ground-truth properties by \Certora and the properties generated by \tool, with a third author breaking ties in case of disagreement.
We welcome other researchers to conduct replication and verification using our released data, available at \website.

\Cref{tab:applicability} shows that \tool can generate comprehensive properties with relatively high recall and reasonable precision.
Most properties generated (26/42) are true positives, and most ground truth properties (8/10) can be successfully reproduced, achieving a satisfactory recall (0.80), reasonable precision (0.64), and fairly good F1-score (0.71).
Delving into project-specific results,
\tool was able to reproduce all the ground-truth properties for four projects including \code{aave\_proof\_of\_reserve}, \code{celo\_governance}, \code{ousd}, and \code{sushi\_benttobox}.
In contrast, \tool reproduced only two ground-truth properties for \code{openzepplin}, suffering the lowest recall and F1-score, although with the highest precision.
We investigated the results and found this is largely because OpenZeppelin~\cite{OpenZeppelin} is a foundational contract library that has been directly imported by nearly all real-world applications, and client code is unlikely to re-implement similar functionality, thus leading to the scarcity of reliable reference properties.
In terms of precision, \code{opyn\_gamma\_protocol} achieves the lowest, reaching only 0.47.
We investigated all 16 false positives about it and later recognized that 11 of these false positives are properties that hold for smart contracts but are not documented in the ground truth by \Certora.



\begin{tcolorbox}[size=title, opacityfill=0.1, breakable]
\textbf{Answer to RQ1}:
\tool can generate comprehensive and high-quality properties, covering 80\% equivalent properties in the ground truth as judged by human experts.
Moreover, the additional properties (FPs) produced by \name generally also hold, complementing the human-written ones.
\end{tcolorbox}

\subsection{RQ2: Vulnerability Detection}
\label{sec:RQ2}

We investigate the applicability of \tool in the vulnerability detection task on well-known smart contract CVEs and the attack incident projects studied by SmartInv.
Note that to mimic the situation of real-world deployment, we set the top-K to top-2, as measured by \mysec\ref{sec:RQ3}, as the best configuration starting from this section.

\noindent
\textbf{CVEs.}
\Cref{tab:cve} demonstrates \tool's effectiveness in detecting 13 smart contract CVEs.
We compared \tool with GPTScan~\cite{sun2024gptscan}, which employs the variable recognition ability of LLMs to instantiate high-level detection patterns for logic bugs, and Slither~\cite{feist2019slither}, a popular static analysis tool used to detect a wide range of common vulnerability types.
In particular, since the original GPTScan covers only ten types of logic bugs, we have enhanced it with the recent unsupervised paradigm~\cite{LLM4Vuln24} and refer to the enhanced version as GPTScan+.
Additionally, Manticore~\cite{mossberg2019manticore} and Mythril~\cite{Mythril} are two bytecode-level symbolic execution tools that automate comprehensive program state exploration and exploit generation of smart contract vulnerabilities.
In \Cref{tab:cve}, the first two columns list CVE names and their vulnerability types, while the remaining columns show the detection results by each tool.

\begin{table}[t]
	\caption{Vulnerability detection results for 13 CVEs.}

	\resizebox{\columnwidth}{!}{
		\begin{tabular}{lccccccr}\toprule
	CVE ID &\rotatebox{75}{Description} &\rotatebox{75}{Avg. Code Sim.} &\rotatebox{75}{\tool} &\rotatebox{75}{GPTScan+} &\rotatebox{75}{Slither} &\rotatebox{75}{Manticore} &\rotatebox{75}{Mythril} \\\midrule
2021-34273 & access control & 0.693 & \y & \y & \x & \x & \x  \\
2021-33403 & overflow & 0.666 & \y & \x & \x & \x & \y  \\
2018-18425 & logic error & 0.704 & \y & \x & \x & \x & \x  \\
2021-3004 & logic error & 0.691 & \x & \x & \x & \x & \x  \\
2018-14085 & delegatecall & 0.662 & \x & \x & \y & \x & \x  \\
2018-14089 & logic error & 0.661 & \y & \y & \x & \x & \x  \\
2018-17111 & access control & 0.636 & \x & \x & \x & \x & \x  \\
2018-17987 & bad randomness & 0.660 & \x & \y & \x & \x & \x  \\
2019-15079 & access control & 0.701 & \y & \x & \x & \x & \x  \\
2023-26488 & logic error & 0.682 & \y & \x & \x & \x & \x  \\
2021-34272 & access control & 0.693 & \y & \y & \x & \x & \x  \\
2021-34270 & overflow & 0.671 & \y & \y & \x & \x & \y  \\
2018-14087 & overflow & 0.661 & \y & \x & \x & \x & \y  \\
	\bottomrule
		\end{tabular}
	}
	\label{tab:cve}
\end{table}

\begin{table}[t]
	\small
	\caption{Evaluation results for 24 attack incident projects from the curated SmartInv benchmark.}
\resizebox{\columnwidth}{!}{
	\begin{tabular}{lrrR{1.5cm}R{1.5cm}R{1.5cm}}\toprule
	Contracts &{Detection} &{\#Property} & Avg. Code Similarity  &{Generation (seconds)} &{Verification (seconds)} \\\midrule
dfxFinance  &\y &8 &0.675 &235 &7 \\
AnySwap  &\x &11 &0.692 &518 &7 \\
Dodo  &\y &17 &0.703 &1,182 &19 \\
Bancor  &\y &19 &0.699 &1,948 &9 \\
BeautyChain  &\y &5 &0.676 &104 &9 \\
Melo  &\y &9 &0.732 &252 &8 \\
BGLD  &\x &9 &0.654 &229 &39 \\
GYMNetwork  &\y &21 &0.681 &274 &71 \\
elasticSwap  &\y &37 &0.681 &1,136 &120 \\
EulerFinance  &\x &23 &0.669 &376 &43 \\
monoSwap  &\y &5 &0.722 &69 &12 \\
nimBus  &\y &32 &0.678 &4,288 &30 \\
VTF &\x &8 &0.672 &358 &21 \\
Nomad &\y &14 &0.673 &590 &70 \\
Umbrella  &\y &14 &0.688 &404 &25 \\
Fortress Loan  &\y &2 &0.668 &71 &5 \\
ShadowFinance  &\y &25 &0.683 &551 &80 \\
Revest  &\y &4 &0.646 &75 &10 \\
Cartel&\y &11 &0.683 &401 &20 \\
sushiSwap  &\x &10 &0.686 &419 &20 \\
ChainSwap  &\x &9 &0.690 &307 &25 \\
Ragnarok  &\y &42 &0.684 &1,890 &88 \\
templeDao  &\y &13 &0.677 &302 &30 \\
BabySwap  &\x &33 &0.679 &1,842 &50 \\
\midrule
Overall &17  &16 &0.683 &743 &34 \\
	\bottomrule
	\end{tabular}
}
	\label{tab:smartInv}
\end{table}

The detection results presented in \Cref{tab:cve} illustrate that \tool outperforms all the comparison tools by detecting 9 out of 13 CVEs, followed by GPTScan detecting five CVEs, Slither detecting only one delegatecall-related CVE, Mythril detecting three overflow-related CVEs, and Manticore detecting zero CVEs.
We also investigated the remaining four CVEs that \tool failed to detect.
It is unknown what valid properties can express the expectation of proper randomness and delegatecall use.
CVE-2018-17111 is caused by the misuse of access control rather than the lack of access control, which is quite challenging for \tool to recognize this subtle difference during property generation.

The ability of \tool can be enhanced by the introduction of newly confirmed vulnerable code and properties into our knowledge database.
As shown in \Cref{tab:property}, the studied properties written by \Certora experts seem to lack support for access control, which could limit the effectiveness of \tool in detecting other wild access control vulnerabilities, even though we realized that \tool has demonstrated a certain level of generalization capability in the aforementioned CVE detection results.

\noindent
\textbf{Attack Incidents.}
\Cref{tab:smartInv} shows the evaluation results on 24 attack incident projects from the curated SmartInv benchmark.
Because the authors of SmartInv did not share their instrumented buggy contract code, and the ground truth and raw experimental results about their generated invariant properties are also missing,
we perform a qualitative rather than a quantitative comparison with SmartInv and will discuss this in \mysec\ref{related}.
In~\Cref{tab:smartInv}, the first two columns list project names and the amount of attack loss.
The remaining columns show the detection results, the number of properties generated, the time used for property generation, and formal verification, respectively.

\tool successfully identified vulnerabilities in 17 out of 24 real-world attack incidents, on average generating 16 properties per project, spending around 12 minutes for property generation and only 34 seconds for formal verification.
For the remaining seven projects that \tool failed to detect, we studied the root causes of their reported vulnerabilities.
We recognized that \tool does not support the runtime context of smart contracts, which may be essential for generating properties in particular use scenarios, for example, deflationary token abuse for the BGLD project, which we leave as future work.

\begin{tcolorbox}[size=title, opacityfill=0.1, breakable]
	\textbf{Answer to RQ2}:
\tool can effectively detect vulnerabilities in both simple and complex smart contracts.
Specifically, \tool has distinguished itself from the current state-of-the-art by detecting 9 out of 13 CVEs.
Additionally, \tool achieved relatively good results in identifying logic bugs in 17 out of 24 attack incidents. 
\end{tcolorbox}

\subsection{RQ3: Generalizability Measurement}
\label{sec:RQ0}

Following the effective vulnerability detection demonstrated in \mysec\ref{sec:RQ2}, we are still interested in whether \name has sufficient generalizability to enable powerful transfer learning.
Given that \name's vector database was constructed from \Certora audit projects, we use the dataset of CVEs and attack incidents used in RQ2 to measure the generalizability of \name's transfer learning during retrieval-augmented property generation for an entirely different dataset.

We use the similarity between the tested code and the retrieved reference code to \textit{indirectly} measure the generalizability of \name's property generation for the vulnerability detection results shown in Table~\ref{tab:cve} and \ref{tab:smartInv}.
That is, if the similarity is lower, that indicates \name's generalizability is stronger.
For the metric of similarity, to avoid bias, we use the average of two popular similarity metrics, namely cosine similarity and Word Mover's Distance~\cite{yi2022fse}.
The results are then presented as standalone columns in Table~\ref{tab:cve} and \ref{tab:smartInv}, respectively.

Overall, we find that the mean average code similarity for all 26 successful cases (i.e., \name successfully generated the correct property) is 0.68, with a range between 0.64 and 0.73, while for all 11 failed cases (i.e., \name failed to generate the appropriate property or the verification was unsuccessful) it is 0.67, with a range between 0.63 and 0.69.
This result has two implications.
First, the absolute similarity values are within a reasonable range—not too high (meaning the test code is very similar to the reference code) nor too low (meaning the vector database fails to provide an effective reference case), indicating that \name demonstrates sufficient generalizability in analyzing an entirely different dataset.
Second, the average similarity for the failed cases is only slightly lower than for the successful cases, 0.67 vs. 0.68, which suggests that the failures were not due to \name lacking the generalizability to test a new piece of code.

\begin{tcolorbox}[size=title, opacityfill=0.1, breakable]
\textbf{Answer to RQ3}:
\name demonstrates sufficient generalizability in analyzing an entirely different dataset through our indirect measurement of the similarity between the tested code and the retrieved reference code.
\end{tcolorbox}

\subsection{RQ4: Influencing Factors}
\label{sec:RQ3}

\begin{figure}[t]
	\includegraphics[width=.98\columnwidth]{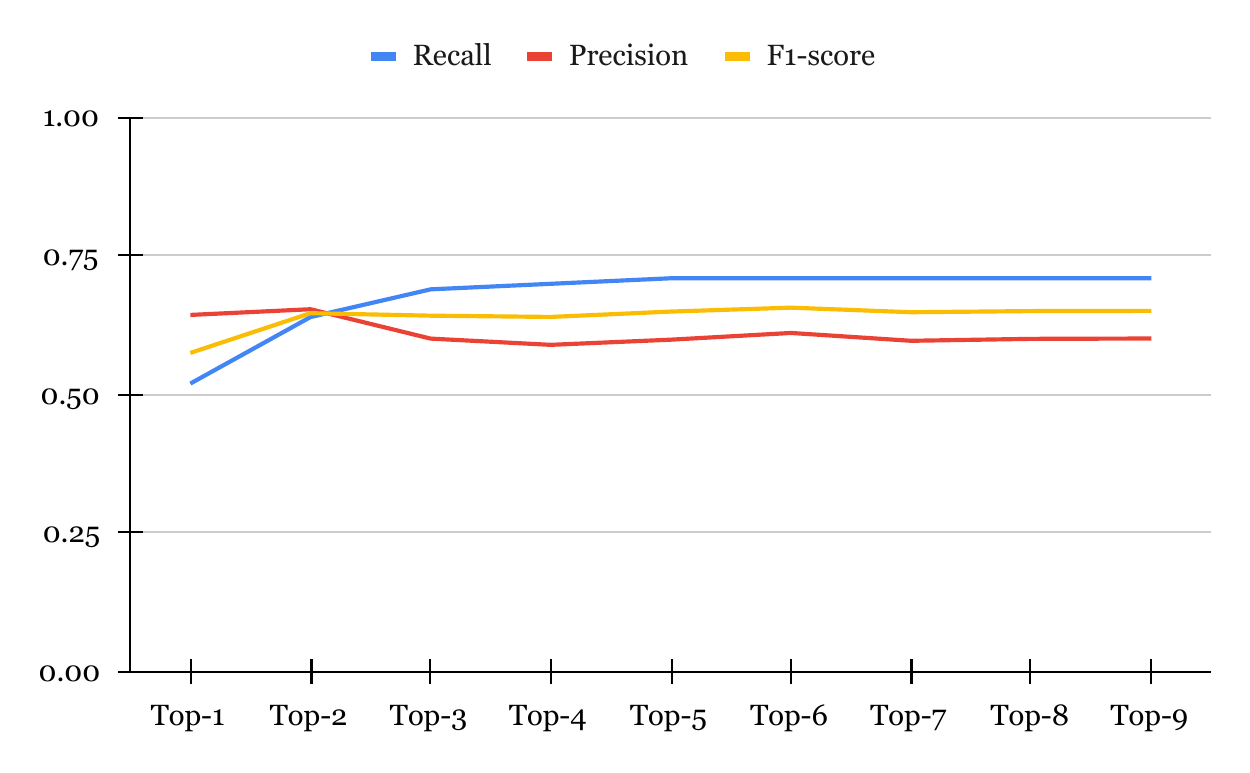}
	\caption{The impact of Top-K settings on property accuracy.}
	\label{fig:topk}
\end{figure}

In our ablation study, we first systematically explored the impact of varying Top-K settings on the property selection process.
Conducting different trials on the same \Certora projects in RQ1, in~\myfig\ref{fig:topk}, we plotted recall, precision, and F1-score for the resulting properties accordingly.
It is clear that all the metrics are above 0.5.
More importantly, when moving from top-1 to top-2, all the metrics increase, although precision has only a very slight increase from top-1 (0.64) to top-2 (0.65),
and afterward, precision goes down and finally fluctuates around 0.60.
Therefore, for the sake of efficiency, we suggest selecting top-2 properties generated for use by external community experts or other compatible program analysis tools.

\begin{table}[t]
	\small
	\caption{The success rate of property generation.}
	\begin{tabular}{lrrr}\toprule
		Method &\#Compilable &\#Failed &Success Rate \\\midrule
		GPT-4.0-turbo w/o fix &234 &136 &0.63 \\
        \tool w/ \mysec\ref{sec:revise} &321 &49 &0.87 \\
		\bottomrule
	\end{tabular}
	\label{tab:successrate}
\end{table}

\begin{figure}[t]
	\centering
	\includegraphics[width=.98\columnwidth]{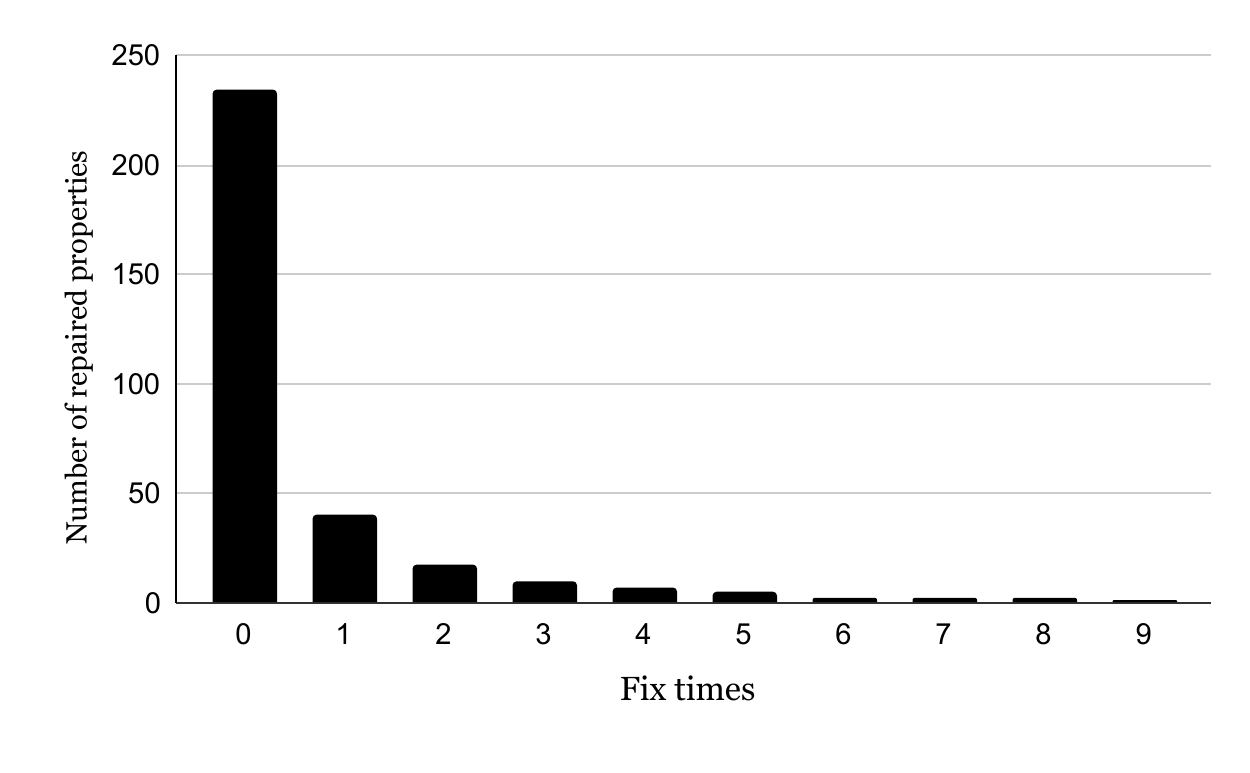}
	\caption{The distribution of property fix times.}
	\label{fig:fix}
\end{figure}

In addition,
we delved into our property generation process with a focus on the success rate of property generation and property repair times using compiler feedback information.
\Cref{tab:successrate} shows that GPT-4.0-turbo without revising or repair achieves a 63\% success rate, which is quite lower compared to \tool (87\%).
\myfig\ref{fig:fix} visualizes the distribution of property repair times, where we capped repair time at nine.
We can see that most compilable properties (84\%) can be generated with no more than five fix attempts.
We also investigated the remaining 49 properties that could not be fixed by \tool and discovered the main compiler error message to be the use of undeclared variables, which may be addressed with a pattern-based approach~\cite{koyuncu2020fixminer}.

\begin{tcolorbox}[size=title, opacityfill=0.1, breakable]
	\textbf{Answer to RQ4}:
\tool can effectively generate compilable properties, with 84\% of all properties being successfully revised within five attempts, and the highest success rate reaching 87\%.
Among them, the top-2 properties achieve the best balance between precision and recall.
\end{tcolorbox}

\begin{figure}[t]
	\centering
	\small
	\begin{minted}[escapeinside=||,texcomments, linenos, breaklines,highlightlines={}]{solidity}
pragma solidity ^0.8.0;
contract SimplifiedStandaloneZkLink {
 address private _owner;
 mapping(address => bool) private _validators;
 uint256 public totalValidatorForwardFee;
 uint256 public totalValidatorForwardFeeWithdrawn;
		
 function withdrawForwardFee(uint256 _amount) external nonReentrant onlyValidator {
   require(_amount > 0, "Invalid amount");
   uint256 newWithdrawnFee = totalValidatorForwardFeeWithdrawn + _amount; |\label{line:start}|
   require(totalValidatorForwardFee >= newWithdrawnFee, "Withdraw exceed"); |\label{line:check}|
			
   totalValidatorForwardFeeWithdrawn = newWithdrawnFee; |\label{line:end}|
   (bool success, ) = msg.sender.call{value: _amount}("");
    require(success, "Withdraw failed");
    emit WithdrawForwardFee(_amount);
  }
}
	\end{minted}
	\caption{The vulnerable withdrawForwardFee function.}
	\label{fig:withdrawForwardFee}
\end{figure}

\begin{figure}[t]
	\small
		\begin{minted}[escapeinside=||,texcomments, linenos, breaklines,highlightlines={}]{solidity}
function withdrawForwardFee(uint256 _amount)
precondition {
 _validators[msg.sender] == true;| \label{line:precond:start}|
 _amount > 0;
 old(totalValidatorForwardFee) >= old(totalValidatorForwardFeeWithdrawn) + _amount; |\label{line:precond:end}|
}
postcondition {
 totalValidatorForwardFeeWithdrawn == old(totalValidatorForwardFeeWithdrawn) + _amount;
 totalValidatorForwardFee - totalValidatorForwardFeeWithdrawn == old(totalValidatorForwardFee) - old(totalValidatorForwardFeeWithdrawn) - amount; |\label{line:assert}|
}
    \end{minted}
    \caption{The property generated for the case in \myfig~\ref{fig:withdrawForwardFee}.}
	\label{fig:pslproperty1}
\end{figure}

\begin{figure}[t]
	\centering
	\small
	\begin{minted}[escapeinside=||,texcomments, linenos, breaklines,highlightlines={}]{solidity}
function addEnvelope(
string calldata envelopeID, bytes32 hashedMerkleRoot,
uint32 bitarraySize, address erc721ContractAddress,
uint256[] calldata tokenIDs
) public {
 require(tokenIDs.length > 0, "Trying to create an empty envelope!");
 MerkleEnvelopeERC721 storage envelope = idToEnvelopes[envelopeID];|\label{line:bug}|  // bug: overwrite storage. 
 envelope.creator = msg.sender;
 envelope.unclaimedPasswords = hashedMerkleRoot;
 envelope.isPasswordClaimed = new uint8[](bitarraySize / 8 + 1);
 envelope.tokenAddress = erc721ContractAddress;
 envelope.tokenIDs = tokenIDs;
	...
}
	\end{minted}
	\caption{The vulnerable addEnvelope function.}
	\label{fig:addEnvelope}
\end{figure}

\begin{figure}[t]
	\small
	\begin{minted}[escapeinside=||,texcomments, linenos, breaklines, breakanywhere, highlightlines={}]{solidity}
rule checkAddEnvelopeCorrectSenderAndCreator() {
 assume(msg.sender == 0x0000000000000000000000000000000000000001);
 string memory envelopeID = "uniqueID"; |\label{line:construct:start}|
 bytes32 hashedMerkleRoot = 0x1234567890abcdef1234567890abcdef1234567890abcdef1234567890abcdef;
 uint32 bitarraySize = 128;
 address erc721ContractAddress = 0x0000000000000000000000000000000000000002;
 uint256[] memory tokenIDs = new uint256[](1);
 tokenIDs[0] = 12345;  |\label{line:construct:end}|
	
 MerkleEnvelopeERC721 storage $envelopeBefore = idToEnvelopes[envelopeID];
 bool $existsBefore = ($envelopeBefore.creator != address(0));
	
 addEnvelope(envelopeID, hashedMerkleRoot, bitarraySize, erc721ContractAddress, tokenIDs); |\label{line:call}|
	
 MerkleEnvelopeERC721 storage $envelopeAfter = idToEnvelopes[envelopeID];
 bool $correctlyAdded = ($envelopeAfter.creator == msg.sender);
 bool $notExistsBefore = ! $existsBefore;
	
 assert($correctlyAdded && $notExistsBefore); |\label{line:exist}|
}
	\end{minted}
    \caption{The property generated for the case in \myfig~\ref{fig:addEnvelope}.}
	\label{fig:pslproperty2}
\end{figure}

\subsection{RQ5: Real-world Impact}
\label{sec:RQ4}

To demonstrate \name's ability to identify zero-day vulnerabilities, we ran \tool on real-world bounty projects hosted by popular platforms such as Secure3~\cite{Secure3} and Code4Rena~\cite{Code4rena}.
\tool successfully generated 22 bug findings for 4 projects, 12 of which have been both confirmed and fixed.
In return, we received \$8,256 in bug bounties from vendors.
In this section, for case studies, we list two zero-day bugs that have been fixed for responsible disclosure, and we do not mention their project source to respect the anonymity policy.

\myfig\ref{fig:withdrawForwardFee} shows that the \code{withdrawForwardFee} function contains a critical vulnerability allowing validators to withdraw more than their allocated share of forwarding fees, potentially leading to unfair distributions and loss of funds.
The vulnerability arises because the function fails to track and limit individual validators' withdrawals according to their proportionate share.
It calculates the new total withdrawn fee by simply adding the requested withdrawal amount \code{\_amount} to \code{totalValidatorForwardFeeWithdrawn} (Lines \ref{line:start}-\ref{line:end}), without considering the requesting validator's entitled share.
The only check performed is against the total collected forwarding fees, ensuring that the new total withdrawn does not exceed this amount (Line~\ref{line:check}).
However, this does not prevent individual validators from withdrawing more than their share.

\tool detected this vulnerability through the generation and verification of function \prepost listed in~\myfig\ref{fig:pslproperty1}.
The pre-conditions(Lines~\ref{line:precond:start}-\ref{line:precond:end}) hold for this contract as they precisely capture the constraints \code{onlyModifier} and the other two \code{require} statements.
The post-conditions describe the expected functionality.
However, one of the post-conditions (Line~\ref{line:assert}), \mintinline[breaklines,breakanywhere]{solidity}{totalValidatorForwardFee - totalValidatorForwardFeeWithdrawn == old(totalValidatorForwardFee) - old(totalValidatorForwardFeeWithdrawn) - amount}, does not hold, thus identifying the contract vulnerability in~\myfig\ref{fig:withdrawForwardFee}.

\myfig\ref{fig:addEnvelope} shows a vulnerable \code{addEnvelope} function where it does not enforce uniqueness of envelope~(Line~\ref{line:bug}), where existing storage can be overwritten arbitrarily.
\myfig\ref{fig:pslproperty2} presents the property generated by~\tool to detect such issue.
Interestingly, \tool can skillfully construct varied input data (Line~\ref{line:construct:start}-\ref{line:construct:end}).
When the function call~(Line~\ref{line:call}) succeeds,
we check the condition (Line~\ref{line:exist}) that same-id envelope does not exist before and the envelope creator equals to the current caller.
In other words, this condition disallows calling \code{addEnvelope} function with same envelope id and ensures the effect of \code{addEnvelope}  will set envelope creator to be the function caller.
Due to the overwrite bug in~\myfig\ref{fig:addEnvelope}, this assertion does not hold for this function.
%
%


%

%
%
%

\subsection{Threats to Validity}
\noindent
\textbf{Internal Validity.}
We evaluated the effectiveness of~\tool on an established \Certora property dataset. 
Nevertheless, there is lack of equivalence checking tool between \Certora-style properties and our proposed PSL-style properties generated by~\tool.
To mitigate this issue, three authors independently reviewed these properties to determine equivalence and we release all the properties generated and the according labeling results for public use.

\noindent
\textbf{External Validity.}
Our findings in vulnerability detection may not apply to other kinds of smart contracts and other types of contract vulnerabilities.
In this work, 
we evaluated~\tool on 13 representative smart contract CVEs covering many kinds of vulnerabilities and 24 real-world victim projects of different application domains.
Moreover, we generated 24 bug findings for high-profile projects to audit and 12 have been confirmed and fixed, with \$8,256 bounty reward.
Therefore, \tool offers a practical formal verification technique for detecting a broader range of smart contract vulnerabilities.

\section{Related Work}
\label{related}


\noindent
\textbf{Vulnerability Detection.}
Numerous automated and semi-automated analysis tools have been proposed to detect smart contract vulnerabilities.
On the one hand, static analysis tools analyze code sequences along the abstract syntax tree of contracts~\cite{feist2019slither} or use a fact-based transformation and query system~\cite{brent2020ethainter,tsankov2018securify} to flag weaknesses and vulnerabilities against expert-written patterns.
In contrast, with test oracles, fuzzers examine runtime behaviors including operation traces~\cite{2018contractfuzzer,nguyen2020sfuzz} and execution effects~\cite{Wang2020OSD,Liu2022FPB} for exploit generation, usually leading to higher precision but lower recall compared with static analyses.
On the other hand, formal verification has been widely employed in techniques to ensure smart contract correctness.
Automated tools like Manticore~\cite{mossberg2019manticore} and Mythril~\cite{Mythril} use symbolic execution to explore as many program states as possible to identify vulnerable behavior with a set of predefined detection rules.
Semi-automated tools require users to provide specification properties, including invariants~\cite{echidna}, function pre-/post-conditions~\cite{wang2018formal}, temporal properties~\cite{stephens2021smartpulse, permenev2020verx}, and other customized rules~\cite{Certora, hildenbrandt2018kevm}.

\tool distinguishes itself by automating property generation using a large language model and proposing a powerful prover based on source code-level symbolic execution of smart contracts, supporting the detection of a wide range of contract vulnerabilities.


\noindent
\textbf{Property Generation.}
Static inference~\cite{tan2022soltype, wang2018formal} and dynamic inference~\cite{Liu2022Invcon, liu2024automated} have been applied in property generation for smart contracts, and recently, machine-learning-based models have also been used for invariant property generation~\cite{liu2022learning, wang2024smartinv}.
VeriSol~\cite{wang2018formal} applies the Houdini algorithm~\cite{flanagan2001houdini} to reason about correct invariant properties from a set of hypothesized candidates.
SolType~\cite{tan2022soltype} discovers type invariants for Solidity smart contracts, requiring developers to add refinement type annotations to the contracts.
However, their properties are limited to arithmetic operations to secure smart contracts from integer overflow and underflow.
InvCon~\cite{Liu2022Invcon} and its subsequent work~\cite{liu2024automated} apply dynamic invariant detection and static inference to produce contract invariants and function \prepost, but they necessitate contracts having sufficient transaction histories.

Our work aligns with previous efforts in machine-learning-based approaches, i.e., Cider~\cite{liu2022learning} and SmartInv~\cite{wang2024smartinv}.
Cider uses a deep reinforcement learning approach but only generates \emph{likely} invariant properties, while \tool can verify all the properties generated with a prover.
Both SmartInv and \tool are powered by large language models.
\tool differs from SmartInv in that we use in-context learning rather than fine-tuning, and our properties generated extend beyond function \prepost.
 
%

\noindent\textbf{LLM-based Security Systems.}
By combining LLMs, various security tasks have been addressed more effectively.
Sun~\etal~\cite{sun2024gptscan} proposed GPTScan, and Li~\etal~\cite{LLift} introduced methods that combine LLMs with static program analysis for vulnerability detection, covering more types of vulnerabilities than traditional tools.
Beyond vulnerability detection, LLMs have been used for other security tasks.
Deng~\etal~\cite{TitanFuzz} proposed TitanFuzz, which utilizes LLMs to guide the fuzzing of deep learning libraries such as PyTorch and TensorFlow.
They also introduced FuzzGPT~\cite{FuzzGPT} to synthesize unusual programs for fuzzing vulnerabilities.
ChatAFL~\cite{ChatAFL} utilizes LLMs to guide the fuzzing of protocols by interpreting the protocol documents. 
Additionally, LLMs have been applied to program repairing tasks, such as ACFix~\cite{zhang2024acfix} and ChatRepair~\cite{Xia_Zhang_2023}.

\section{Conclusion}
\label{conclusion}
In this paper, we proposed retrieval-augmented property generation for smart contracts by utilizing LLMs' in-context learning capabilities.
We implemented this approach in a tool called \tool and addressed challenges to ensure the generated properties are compilable, appropriate, and runtime-verifiable.
Our evaluation results indicate that \tool can detect many real-world contract vulnerabilities, especially in high-profile projects, collectively receiving \$8,256 in bounty rewards from vendors.
For future work, we plan to include more comprehensive contract context information, such as documentation, in our approach and enhance \tool with richer property knowledge from various sources.

\section*{Acknowledgement}
We thank all the reviewers for their constructive feedback on this paper.
This research/project is supported by the Singapore Ministry of Education Academic Research Fund Tier 1
(RG12/23), the Nanyang Technological University Centre for Computational Technologies in Finance
(NTU-CCTF), the National Research Foundation Singapore and DSO National Laboratories under the AI Singapore Programme (AISG Award No: AISG2-RP-2020019), and the National Research Foundation, Prime Minister's Office, Singapore under its Campus for Research Excellence and Technological Enterprise (CREATE) programme.
Any opinions, findings and conclusions or recommendations expressed in this material are those of the author(s) and do not reflect the views of NTU-CCTF, National Research Foundation, Singapore and Cyber Security Agency of Singapore.
\bibliography{main}
\bibliographystyle{IEEEtranS}

\appendix

\subsection{Supplementary Material}
\label{sec:appendixA}

Table~\ref{tab:appendix} lists the raw information for all the {61} Certora projects, of which we collected 23 projects with available code and properties.

\begin{table*}[!h]
    \caption{The raw information for all the {61} \Certora projects.}
    \large
    \centering
\scalebox{.68}{
	\begin{tabular}{lrrrrr}\toprule
	Report Name &Year &Month &Included &\#Property \\\midrule
	Aave CLSynchronicity Price Adapter &2022 &December &\x & \\
	Aave GHO Stablecoin &2023 &March &\y &35 \\
	Aave Governance V2 Update &2022 &September &\x & \\
	Aave L2 Bridge &2022 &July &\y &42 \\
	Aave Proof of Reserve &2022 &November &\y &3 \\
	Aave Protocol V2 &2020 &December &\y &17 \\
	Aave Rescue Mission Phase 1 &2023 &January &\y &1 \\
	Aave Staked Token v1.5 &2023 &February &\y &11 \\
	Aave Static aToken &2023 &April &\y &24 \\
	AAVE Token V3 &2022 &September &\x & \\
	Aave V2 AStETH &2022 &August &\x & \\
	Aave V3 &2022 &January &\y &59 \\
	Aave V3 BTC.b Listing Steward &2022 &September &\x & \\
	Aave V3 MAI \& FRAX Listing Stewards &2022 &August &\x & \\
	Aave V3 PR \#820 &2023 &March &\x & \\
	Aave V3 sAVAX Listing Steward &2022 &July &\x & \\
	Aave V3 sUSD Listing Steward &2022 &August &\x & \\
	Aave V3.0.1 &2022 &December &\x & \\
	Aave Vault &2023 &June &\y &16 \\
	Aave-StarkNet L1-L2 Bridge &2022 &October &\y &10 \\
	Balancer &2022 &September &\x & \\
	Balancer V2 &2021 &April &\x & \\
	Balancer V2 (Issues only) &2021 &April &\x & \\
	Balancer’s Timelock Authorizer Verification Report &2023 &May &\x & \\
	Benqi’s Liquid Staking Contracts &2022 &April &\x & \\
	Celo Core Contracts Release 4 &2021 &May &\x & \\
	Celo Governance Protocol &2020 &May &\y &35 \\
	Compound V1 Price Oracle &2018 &September &\x & \\
	Compound V3 Comet &2022 &July &\x & \\
	Compound’s MoneyMarket v2 formal verification report &2019 &August &\y &41 \\
	Compound’s Open-Oracle with Uniswap Anchor &2020 &August &\x & \\
	Daoism &2022 &October &\x & \\
	dcSpark &2022 &December &\x & \\
	dForce Lending Protocol &2021 &February &\x & \\
	Euler &2021 &November &\x & \\
	Furucombo &2021 &May &\y &20 \\
	Kashi Lending Protocol &2021 &March &\x & \\
	Keep’s Fully-backed bonding contract &2020 &November &\y &13 \\
	Lido V2 &2023 &April &\y &1 \\
	Lyra &2021 &May &\x & \\
	Master Chef V2 &2021 &April &\x & \\
	Notional Finance V2 &2021 &November &\y &30 \\
	OOPSLA'2020 &2020 &- &\x & \\
	Open Zeppelin &2022 &April &\y &80 \\
	Open Zeppelin &2022 &June &\x & \\
	OpenZeppelin Governance contracts &2021 &December &\x & \\
	Opyn Gamma Protocol &2020 &December &\y &33 \\
	Orchid’s Smart Contracts &2019 &December &\x & \\
	Origin OUSD Token &2021 &February &\y &16 \\
	Popsicle V3 Optimizer &2021 &November &\y &21 \\
	Radicle Drips &2023 &January &\y &36 \\
	Rolla Finance &2021 &August &\x & \\
	SaaS Verification Report by Blockswap Labs &2022 &July &\x & \\
	SaaS Verification Report by Silo &2022 &July &\x & \\
	Sushi BentoBox &2021 &February &\y &22 \\
	Sushi Compound Strategy &2021 &April &\x & \\
	SushiSwap ConstantProductPool &2021 &November &\x & \\
	SushiSwap TridentRouter &2021 &November &\x & \\
	Synthetix Multi-Collateral Loans &2020 &December &\x & \\
	Trader Joe &2022 &March &\y &98 \\
	Zesty &2021 &July &\x & \\
	\bottomrule
	\end{tabular}
}
	\label{tab:appendix}
\end{table*}

\subsection{An Example of CVL}
\label{sec:CVLexample}

\begin{figure}[h]
	\small
	\begin{minted}[escapeinside=||,texcomments, linenos, breaklines, breakanywhere, highlightlines={}]{solidity}
hook Sstore _checkpoints[KEY address account][INDEX uint32 index].votes uint224 newVotes (uint224 oldVotes) STORAGE {    
 havoc userVotes assuming
   userVotes@new(account) == newVotes;

 havoc totalVotes assuming
   totalVotes@new() == totalVotes@old() + newVotes - userVotes(account);
		
  havoc lastIndex assuming
    lastIndex@new(account) == index;
}	
\end{minted}
\caption{Part of the CVL specification for ERC20Votes.}
	\label{fig:cvl}
\end{figure}

Following the description in \mysec\ref{sec:lang}, we provide a CVL example here to illustrate its reliance on a low-level rather than a high-level execution model for smart contracts.

\myfig\ref{fig:cvl} shows part of the CVL specification for ERC20Votes\footnote{\url{https://github.com/Pr0pertyGPT/PropertyGPT/blob/main/certora_projects/openzepplin/specs/ERC20Votes.spec}}.
It specifies the semantic behavior for updating the field item \texttt{votes} of \texttt{mapping(address => Checkpoint[]) private \_checkpoints}, a data structure used to record user votes.
Essentially, the specification states that after users cast new votes, the recorded user votes should be updated accordingly, the total votes should be recalculated, and the last index, i.e., the size of the user checkpoint array, should also be updated accordingly.

In our experience, writing such specifications is non-trivial for smart contract developers, as \myfig\ref{fig:cvl} requires explicit semantic definitions for each field item of complex data structures.
Additionally, explicitly declaring low-level opcodes such as \texttt{sstore} and specifying the variable type of involved field items (e.g., STORAGE indicating a storage variable) burdens users with subtle and exhaustive details of smart contract data storage.

In comparison, PSL extends Solidity, and its execution follows the well-studied Solidity semantics, allowing users to write specifications without needing to know the intricate details of the verification process.

\subsection{Compilation Phase of PSL Specifications}
\label{sec:compilationDetails}


PSL is developed as a variant of Solidity, where we modify the Solidity compiler by adding new syntactic structures and imposing certain constraints to support the compilation of specifications written in PSL.
Specifically, we add the following keywords by modifying Solidity compiler v0.8.17:
\texttt{invariant} for declaring invariant specification code blocks,
\texttt{rule} for declaring customized rule code blocks,
and \texttt{precondition} and \texttt{postcondition} for function preconditions and postconditions.
We also add the keyword \texttt{assume} for verification purposes (c.f.~\myfig\ref{fig:psl}).
Moreover, we implement checks to permit only expression statements in invariant, precondition, and postcondition specifications.
Through these measures, our customized PSL compiler can accept specifications written in PSL and validate their syntactic correctness.

\subsection{Learning Optimal Coefficients via Training a Linear Regression Model}
\label{sec:Coefficients}

Following the introduction in \mysec\ref{sec:rank}, this section details how we train a linear regression model to learn optimal coefficients, which is vital for ranking the top-k appropriate properties as the final property generation result.

\noindent
\textbf{Data Preparation.}
We used \tool to generate a total of 3,622 property generation records.
Specifically, for each property generation, we randomly selected a known subject code $f$ from the Cetora dataset, and then \tool yielded a rule property $\phi_1$ for it.
We measured $X_{raw}(f, g)$, $X_{summary}(f, g)$, $Y_{raw}(\phi_1, \phi_2)$, and $Y_{summary}(\phi_1, \phi_2)$.
Additionally, we computed the actual score $\hat{Score}(f, \phi_1) = Y_{summary}(\phi_1, \hat{\phi_1})$, where $\hat{\phi_1}$ is the known rule property of $f$ in the Certora dataset.
Our training aims to produce ${Score}(f, \phi_1)$ to approximate $\hat{Score}(f, \phi_1)$ using the four features mentioned in \mysec\ref{sec:rank}.

\noindent
\textbf{Model Training.}
We trained an Ordinary Least Squares (OLS) linear regression model using the above data and evaluated the model's performance using multiple metrics: Mean Absolute Error (MAE), Mean Squared Error (MSE), Root Mean Squared Error (RMSE), Coefficient of Determination ($R^2$), Mean Absolute Percentage Error (MAPE), and Mean Deviation Error (MDE). Finally, we obtained the following weight coefficients and performance results:
\begin{itemize}
		\item Coefficients: $\alpha$: 0.134, $\beta$: 0.556, $\gamma$: 0.141, and $\eta$: 0.168
		\item Performance metrics: MAE: 0.0239, MSE: 0.0008, RMSE: 0.0291, $R^2$: 0.1294, MAPE: 2.7810, and MDE: -0.0022
\end{itemize}
These coefficient settings achieve relatively good performance.
Additionally, we conducted other primitive experiments and found that combining fewer features in the prediction model led to a decline in performance metrics. Therefore, we believe that the selected four-feature combination can rank the generated properties with reasonably high accuracy.

\end{document}